\documentclass{elsart}
\usepackage{graphicx}

\usepackage{amssymb}

\usepackage{lineno}
\usepackage{bm}
\usepackage{pifont}
\usepackage{color}
\usepackage{marvosym}

\begin{document}

\begin{frontmatter}



\title{In-medium chiral $SU(3)$ dynamics and hypernuclear
  structure
}
\author[label2]{P. Finelli}
\ead{paolo.finelli@bo.infn.it},
\author[label3]{N. Kaiser},
\author[label4]{D. Vretenar},
\author[label3]{W. Weise},
\address[label2]{Physics Department, University of Bologna, I-40126 (Italy)}
\address[label3]{Physik-Department, Technische
  Universit\"at M\"unchen,\\ D-85747 Garching (Germany)}
\address[label4]{Physics Department, University of Zagreb, HR-10002 (Croatia)}


\begin{abstract}
A previously introduced relativistic energy density functional, 
successfully applied to ordinary nuclei, is extended to hypernuclei.
The density-dependent mean field and the spin-orbit potential 
are consistently calculated for a $\Lambda$ hyperon in the nucleus
using the $SU(3)$ extension of in-medium chiral perturbation theory.
The leading long range $\Lambda N$ interaction arises from
kaon-exchange and $2\pi$-exchange with $\Sigma$ hyperon
in the intermediate state.
Scalar and vector mean fields reflecting in-medium changes 
of the quark condensates are constrained by QCD sum rules.
The model, applied to oxygen as a test case,
describes spectroscopic data in good agreement with experiment.
In particular, the smallness of the $\Lambda$ spin-orbit interaction
finds a natural explanation in terms of an almost complete
cancellation between scalar-vector background contributions and long-range
terms generated by two-pion exchange.
\end{abstract}


%
%
\journal{Physics Letters B}

\end{frontmatter}


1. {\it Introduction.} Ever since the discovery of the first 
$\Lambda$ hypernucleus in 1953~\cite{danysz},  hypernuclear 
physics has been an active research 
area~\cite{hyper_review_1,hyper_review_2}, 
in particular once spectroscopic investigations
using $(K^-,\pi^-)$ reactions became available~\cite{Hashimoto_rev}.
A most intriguing result has been the extraordinary weakness of 
the $\Lambda$-nucleus spin-orbit interaction.
In recent years high quality $\Lambda$-hypernuclear spectra produced
by $(\pi^+,K^+)$ reactions confirmed this result. For example, 
measurements of $E1$-transitions from $p$- to $s$-shell orbitals
of a $\Lambda$ hyperon in $^{13}_{\Lambda}$C gave a
$p_{3/2}-p_{1/2}$ spin-orbit splitting of 
only $(152 \pm 90)$ keV
~\cite{Kohri:2001ncAjimura:2001naAkikawa:2002tm} 
(much smaller than the
corresponding $6$ MeV in ordinary $p$-shell nuclei).
The same conclusion is drawn from an  
analysis of the excitation energy difference
between the $0^+_1$ and $2^+_1$ states of $^{16}_{\Lambda}$O
\cite{Hashimoto_rev}.
The experimental evidence therefore suggests that 
the $\Lambda$ spin-orbit interaction is indeed very weak.

Early theoretical attempts to describe hypernuclear spectra~\cite{wei_bro} 
assumed weak couplings between $\Lambda$ and exchanged bosons in a relativistic
mean field model, where ``weak'' means that the scalar and vector fields
experienced by the $\Lambda$ in the hypernucleus
have $1/3$ of the strength of the corresponding scalar and vector mean fields 
for the nucleons. In contrast, a quark model description in which 
the strange quark
inside the $\Lambda$ does not interact with the up and down quarks
of the nucleons, suggests a ratio $2/3$ between couplings of the 
$\Lambda$ and the nucleons. 
Subsequently Pirner, Noble and Jennings~\cite{noblepirnerjenn}
reconciled the empirical findings with quark model
predictions by introducing a strong (negative) $\omega \Lambda \Lambda$
tensor coupling which generates
a spin-orbit force with opposite sign so as to yield a small net result.
Phenomenological studies along this line~\cite{tensor} have been useful
in reproducing the empirical single particle levels
for a large set of hypernuclei. However, such a strong (negative)
$\omega \Lambda \Lambda$ tensor coupling was considered unnatural. 

In recent years new approaches emerged in order to
investigate hypernuclear spectra from a microscopic point of view.
A connection between quark model features and a relativistic
one-boson picture was drawn by the Quark-Meson 
Coupling (QMC) model~\cite{Tsushima}. 
Lenske {\it et al.} ~\cite{Keil:1999hk}
used a density dependent relativistic approach
in which the $\Lambda$-meson couplings are partly determined 
from a theoretical $\Lambda$N T-matrix and partly
fitted to a selected set of data.
Despite these constraints, it turned out not to be possible
to understand the anomalously small $\Lambda$-nucleus spin-orbit force.

The relevance of explicit two-pion exchange contributions 
to the nuclear force and low-energy observables has become 
a generally accepted fact.
In Refs.~\cite{Kaiser:2001jx,Fritsch:2004nx} it has been demonstrated that
iterated one-pion exchange and irreducible $2\pi$ exchange processes
with inclusion of Pauli blocking and $\Delta$ intermediate states can
generate the correct nuclear binding in the nuclear matter case.
For hypernuclei the importance of correlated $2\pi$ exchange 
was pointed out in Ref.~\cite{wei_bro}
which emphasized the role of $\Sigma^*$ resonances as intermediate states.
Only recently~\cite{Kaiser:2004fe}
the long-range $\Lambda$N interaction arising from kaon and $2\pi$ exchange, 
with $\Sigma$ hyperons as intermediate states and medium insertions, 
has been explicitly calculated in a controlled expansion in 
powers of the Fermi momentum $k_f$. 

2. $\Lambda$-{\it hypernuclei in the 
context of in-medium chiral} $SU(3)$ {\it  dynamics.}
In this work we present an extension to hypernuclei of a 
relativistic nuclear energy density functional~\cite{Finelli:2003fk,
Finelli:2005ni,Finelli:2006xh}
that combines relevant features of chiral dynamics
and the symmetry breaking pattern of low-energy QCD.
Chiral pionic fluctuations in combination with Pauli 
blocking effects~\cite{Kaiser:2001jx}, $\Delta$ excitations
and three-nucleon ($3N$) interactions~\cite{Fritsch:2004nx} 
are superimposed on the condensate background fields
and produce the nuclear binding. Scalar and vector mean fields
representing the in-medium changes of the quark condensates, 
which cancel almost completely in their sum, 
act coherently in their difference to generate the large spin-orbit potential
for nucleons in nuclei.
This model has been successfully 
applied~\cite{Finelli:2003fk,Finelli:2005ni,Finelli:2006xh}
to the description of ground states and collective 
excited states of open-shell nuclei.
In the present work this approach is generalized
to study $\Lambda$ hypernuclei. In particular, we
examine whether the novel mechanism 
for the suppression of the spin-orbit potential
proposed in Ref.~\cite{Kaiser:2004fe} (and recently also followed in 
Ref.~\cite{Camalich:2006is}) at the nuclear matter level works as well
in finite hypernuclei.


2.1 {\it The model.} To describe hypernuclei we generalize 
the relativistic density functional 
previously introduced (see Sect. 2.2 in Ref.~\cite{Finelli:2005ni})
adding the hyperon contribution:
\begin{equation}
E_0^{~}[{\rho}] = E_0^N[{\rho}] + E_0^\Lambda[{\rho}] \;,
\end{equation}
where $E_0^N[{\rho}]$ describes the core of protons 
and neutrons (see Eq. (12) in
Ref.~\cite{Finelli:2005ni}) and 
$E_0^\Lambda[{\rho}]$ is the leading-order term 
for the single $\Lambda$ hyperon,
decomposed in free and interaction parts:
\begin{equation}
E_0^\Lambda[{\rho}] = E_{\rm free}^\Lambda[{\rho}] 
+ E_{\rm int}^\Lambda[{\rho}] \;,
\end{equation}
with
\begin{eqnarray}
E_{\rm free}^\Lambda & = & 
\int d^3r \langle \phi_0 | \bar{\psi}_\Lambda [-i 
\bm{\gamma} \cdot \bm{\nabla} + M_\Lambda ]
\psi_\Lambda |\phi_0 \rangle \\
E_{\rm int}^{\Lambda} & = & 
\int d^3r \left\{ \langle \phi_0 | 
G^{\Lambda}_S ({\rho}) \left( \bar{\psi} \psi \right) \left(
\bar{\psi}_\Lambda \psi_\Lambda \right) | \phi_0 \rangle + \right. \nonumber\\
& ~ & \left. \quad \quad \quad \langle \phi_0 | G^{\Lambda}_{V} 
({\rho}) \left( \bar{\psi} \gamma_\mu \psi \right) \left(
\bar{\psi}_\Lambda \gamma^\mu \psi_\Lambda \right) | \phi_0 \rangle
\right\} \; .
\end{eqnarray} 
Here $|\phi_0 \rangle$ denotes the (hypernuclear) ground state.
$E_{\rm free}^{\Lambda}$ is the contribution to the energy from
the free relativistic
hyperon including its rest mass $M_\Lambda$. The interaction term
$E_{\rm int}^{\Lambda}$ includes density dependent hyperon-nucleon vector
($G^{\Lambda}_V$) and scalar ($G^{\Lambda}_S$) couplings.
They receive mean-field contributions from in-medium changes of 
the quark condensates (identified with superscript $(0)$) 
and from in-medium kaon- and two-pion exchange processes
(with superscript $(K, \pi)$):
\begin{equation}
G^{\Lambda}_i({\rho}) = G^{\Lambda (0)}_i + G^{\Lambda (K, \pi)}_i({\rho})
\quad {\rm with} \quad i=S,V \; .
\end{equation}
Minimization of the ground-state energy leads to coupled 
relativistic Kohn-Sham equations for 
the core nucleons and the single $\Lambda$ hyperon. 
Using the same notation as 
in Ref.~\cite{Finelli:2005ni} we have:
\begin{eqnarray}\
\label{dir_eq}
\left[ -i \bm{\gamma} \cdot \bm{\nabla} + M_N + \gamma_0 \left( \Sigma_V +
 \Sigma_R + \tau_3 \Sigma_{TV} \right) + \Sigma_S + \tau_3 \Sigma_{TS} \right] 
\psi_k & = & \epsilon_k \psi_k \\
\label{dir_eq_2}
\left[ -i \bm{\gamma} \cdot \bm{\nabla} + M_\Lambda + \gamma_0 \Sigma_V^\Lambda +
 \Sigma_S^\Lambda \right] \psi_\Lambda & = & \epsilon_\Lambda \psi_\Lambda \; ,
\end{eqnarray}
where $\psi_k$ and $\psi_\Lambda$ are now the wave functions of 
the Kohn-Sham single particle orbits for the nucleons and the 
$\Lambda$, respectively. 
These single particle Dirac equations together with the 
self-energies $\Sigma_i$ are
solved self-consistently in the ``no-sea'' approximation~\cite{SerotFurnstahl}.
It is important to note that rearrangement 
contributions $\Sigma_R$~\cite{Fuchs:1995as} 
in the previous equations are confined to the nucleon 
sector because all the density dependent couplings are 
polynomials in $k_f$ (and consequently in fractional 
powers of the baryon density
through the relation $\rho = 2\, k_f^3/(3 \pi^2)$), 
and there is no hyperon Fermi sea.
The $\Lambda$ self-energies are
\begin{equation}
\Sigma_V^\Lambda  = G^{\Lambda}_V(\rho)\, \rho~ ,~~~~~~~~~~~~~~~~
\Sigma_S^\Lambda  =  G^{\Lambda}_S(\rho)\, \rho_S~  ,
\end{equation}
in terms of the nuclear baryon and scalar 
densities, $\rho$ and $\rho_S$. Inclusion and 
analysis of corrections from derivative and tensor 
terms are postponed to a forthcoming paper.

In the following paragraphs we separately analyze 
the different contributions to the density dependent 
$\Lambda$-nuclear couplings
$G^{\Lambda}_i({\rho})$ arising from the kaon- and 
two-pion-exchange induced $\Lambda$-nucleus 
potential, the condensate background mean fields 
and the pionic $\Lambda$-nucleus spin-orbit interaction.

\begin{figure}
\vspace{1cm}
\begin{center}
\includegraphics[scale=0.4,angle=0]{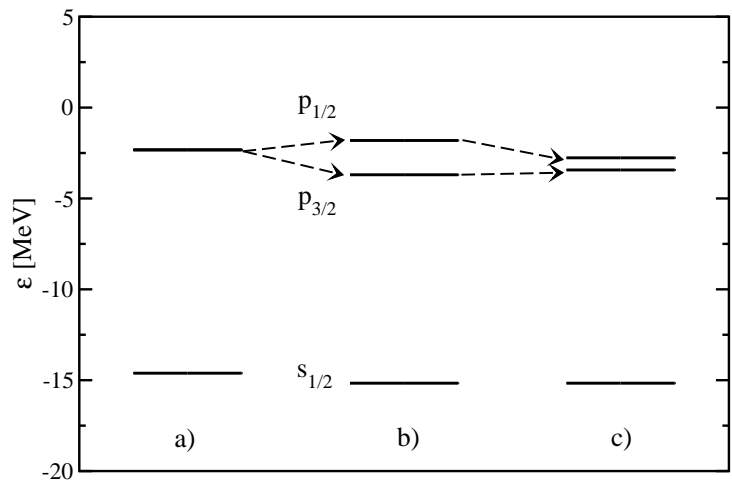}
\caption{\label{figA} 
$\Lambda$ single particle energy levels in $^{16}_\Lambda O$. 
Case $a$: using the single particle potential based on the density dependent coupling strengths including chiral $K$ and $2\pi$ exchange as determined in Ref.~\cite{Kaiser:2004fe}
(see Sect. II). 
Case $b$: adding the spin-orbit effect from in-medium quark condensates, 
with reduction factor $\chi = 2/3$ according to a simple quark model~\cite{noblepirnerjenn}.
Case $c$: additional compensating effect of the chiral SU(3) spin-orbit potential
from the second order $\Lambda N$ tensor force with intermediate $\Sigma$ (see Eq.(\ref{so})).
}
\end{center}
\end{figure}

2.2 {\it Kaon- and two-pion induced mean field.} 
In Ref.~\cite{Kaiser:2004fe} the density dependent 
self-energy for a zero momentum $\Lambda$ hyperon 
in isospin-symmetric nuclear matter has been calculated 
at two-loop order in the energy density. 
This calculation systematically includes kaon-exchange 
Fock terms and two-pion exchange with $\Sigma$ hyperon
and Pauli blocking effects in the intermediate state. 
This self-energy is translated into a mean field potential $U_\Lambda (k_f)$.
A cutoff scale ${\bar{\Lambda}}$ 
(or equivalently, a contact term) represents short distance
(high momentum) dynamics not resolved at scales characteristic 
of the Fermi momentum. Tuning this scale to 
${\bar{\Lambda}} = 0.71$ GeV, remarkably close
to the value $0.7$ GeV used in Ref.~\cite{Kaiser:2004fe},
the depth of the $\Lambda$-nuclear central 
potential is fixed such that the $p$-state 
in $^{16}_\Lambda O$ are close to the empirical values 
($\epsilon_\Lambda^p = -1.86 \pm 0.06$ 
MeV \cite{Hashimoto_rev})\footnote{
all the calculations are carried out for $^{17}_\Lambda$O.
}.
We then follow the procedure outlined in the 
Appendix A of Ref.~\cite{Finelli:2005ni}
and determine the equivalent density dependent 
$\Lambda$ point coupling vertices
$G^{\Lambda (K, \pi)}_S({\rho})$ and $G^{\Lambda (K, \pi)}_V({\rho})$.
For the nucleon sector of the energy density functional we use
the parameter set FKVW~\cite{Finelli:2005ni}: 
four parameters related to contact
terms that appear in the ChPT treatment of nuclear matter, one
parameter for the derivative (surface) term, and two more for
the strengths of the condensate background scalar and vector
mean fields.
In Fig.~\ref{figA} (case {\it a}) the resulting $\Lambda$ single particle 
energy levels of $^{16}_\Lambda O$ are plotted. 
At this stage the {\it p} shell spin-orbit
partners are practically degenerate as expected from
previous investigations~\cite{Finelli:2003fk}. The energies of the 
degenerate doublets are, by construction, close to their observed positions.
Even the calculated energy of the {\it s} state is realistic although 
slightly too large in comparison with the empirical 
$\epsilon_\Lambda^s =-12.42 \pm 0.05$ MeV for $^{16}_\Lambda O$~\cite{Hashimoto_rev}.

Up to this point in-medium chiral SU(3) dynamics 
(with $K$ and $2\pi$ exchange) provides the necessary 
binding of the system but no spin-orbit force. As already
shown in Ref.~\cite{Finelli:2003fk} inclusion of derivative couplings 
does not remove the degeneracy of the spin-orbit doublets.


2.3 {\it Background scalar and vector mean fields.} 
In contrast to the mean field induced by kaon and 
two-pion exchange, condensate background self-energies 
of the $\Lambda$ produce a sizeable spin-orbit potential 
in a way analogous to what has been pointed out  
in Ref.~\cite{Finelli:2003fk,Finelli:2005ni} for the nucleon case.
Under the assumption that only non-strange quarks are
involved in interactions with the background fields, one expects
a reduction of the corresponding couplings,
\begin{equation}
G^{\Lambda (0)}_{S,V} =  \chi \,G^{(0)}_{S,V} ~,
\end{equation}
by a factor $\chi = 2/3$~\cite{noblepirnerjenn}, 
where $G^{(0)}_V$ and $G^{(0)}_S$ are
the vector and the scalar couplings to nucleons, 
arising from in-medium changes
of the quark condensates, $\langle\bar{q}q\rangle$ 
and $\langle q^\dagger q\rangle$.
The $G^{(0)}_V$ and $G^{(0)}_S$ have been determined, 
in good agreement with leading-order QCD sum 
rules estimates~\cite{Cohen:1994wm},
by fitting ground state properties 
of finite nuclei~\cite{Finelli:2005ni}.

For illustration we plot in Fig.~\ref{figA} 
(case {\it b}) the $\Lambda$ single particle 
energy levels with inclusion of these scalar 
and vector mean fields using $\chi = 2/3$. 
Now the {\it p} shell spin-orbit partners are no longer degenerate
and a spin-orbit splitting of about $\sim 2$ MeV results.
The choice $\chi = 2/3$ is, of course, a simplistic estimate.
A detailed  QCD sum rule analysis
suggests a reduction to 
$\chi \sim 0.4 - 0.5$~\cite{Cohen:1994wm,Jin:1993fr}, and to
even smaller values if corrections from in-medium
condensates of higher dimensions are taken into account.
We shall therefore be guided by such reduced 
values of $\chi$. Nonetheless, the $\Lambda$-nuclear 
spin-orbit force is evidently still far too strong 
at this level, just as in the phenomenological 
relativistic "sigma-omega" mean field models.

2.4 {\it $\Lambda$-nuclear spin-orbit interaction from 
chiral SU(3) two-pion exchange.}
In Ref.~\cite{Kaiser:2004fe} the $\Lambda$-nucleus 
spin-orbit interaction generated by the
in-medium two-pion exchange $\Lambda$N interaction, 
has been evaluated as follows. In the spin-dependent 
part of the self-energy of a $\Lambda$ hyperon 
scattering in slightly inhomogeneous nuclear matter 
from initial momentum $\vec{p} - \vec{q}/2$ to final
 momentum $\vec{p} + \vec{q}/2$, one identifies a 
spin-orbit term, $\Sigma_{ls}^\Lambda(k_f) = 
{i\over 2}U_{ls}^\Lambda(k_f)\,
\vec{\sigma}\cdot(\vec{q}\times\vec{p}\,).$ 
It depends only on known $SU(3)$ axial vector 
coupling constants and on the mass difference 
between $\Lambda$ and $\Sigma$. The relevant 
momentum space loop integral is finite and 
hence model independent in the sense that no 
regularizing cutoff is required. 
The result, $U_{ls}^\Lambda(k_f^{(0)})\simeq -15$ MeV fm$^2$ 
at $k_f^{(0)} \simeq 1.36$ fm$^{-1}$, has a sign {\it opposite}
 to the standard nuclear spin-orbit 
interaction\footnote{Recall that nuclear Skyrme phenomenology gives
$U_{ls}^N(k_f^{(0)}) = 3 W_0 \rho_0/2  \simeq 30$ MeV fm$^2$ for 
the nucleonic spin-orbit coupling strength~\cite{Chabanat:1998}.}. 
Evidently, this term tends to largely cancel
the spin-orbit potential generated by the scalar-vector background mean field.

It is important to note that such a ``wrong-sign" 
spin-orbit interaction (generated by the second 
order tensor force from iterated pion exchange) 
exists also for nucleons in ordinary nuclei \cite{KFW:2003}. 
However, this effect is compensated to a large 
extent by the three-body spin-orbit force involving 
virtual $\Delta(1232)$ isobar excitation \cite{Kaiser:2003}, 
so that the spin-orbit interaction from the strong 
scalar-vector mean fields prevails\footnote{The intimate connection 
between large
scalar and vector mean fields of opposite sign in nuclear matter and the
short-range spin-orbit term of realistic NN-potentials has recently been
demonstrated in Ref.~\cite{Plohl:2006hy}
}. For a $\Lambda$ 
hyperon, the analogous three-body effects do not 
exist and the compensation is now between spin-orbit 
terms from the (weaker) scalar-vector mean field and 
the in-medium second order tensor force from iterated 
pion exchange with intermediate $\Sigma$. 
The small $\Sigma\Lambda$ mass splitting, 
$M_\Sigma - M_\Lambda = 77.5$ MeV, plays 
a prominent role in this mechanism.

In order to estimate the impact of 
this genuine ``wrong-sign" $\Lambda$-nuclear spin-orbit term, we introduce
\begin{equation}
\Delta\mathcal{H}_{ls}^{\Lambda} = 
-i\,{U_{ls}^\Lambda(k_f^{(0)})\over 2r}\,{df(r)\over dr}\,
\vec{\sigma}\cdot(\vec{r}\times\vec{\nabla})~, 
\end{equation}
with the normalized nuclear density profile $f(r) = \rho(r)/\rho(r=0)$.
We then evaluate the corrections to the 
$\Lambda$ single particle energies $\epsilon_{\Lambda}$ in first order 
perturbation theory:
\begin{equation} 
\label{so}
\epsilon_{\Lambda}' = \epsilon_{\Lambda} + 
\langle \phi | \Delta\mathcal{H}_{ls}^{\Lambda} | \phi \rangle \;,
\label{eqn:ls}\end{equation}
where $|\phi \rangle$ denotes the self-consistent solution of the
system of Dirac single-baryon equations (\ref{dir_eq})
and (\ref{dir_eq_2}).
In Fig.~\ref{figA} (case {\it c}) one observes that 
the resulting {\it p} shell single particle energy levels, 
corrected according to Eq.(\ref{eqn:ls}), are then close to being degenerate. 
The spin-orbit splitting is now strongly reduced but still 
too large in comparison with empirical estimates.
This is considered a consequence of the possibly too large 
quark model factor $\chi = 2/3$. Reducing this factor to 
the range of values compatible with QCD sum rules, the 
final step towards the empirical, almost vanishing 
spin-orbit splitting for the $\Lambda$ can indeed be accomplished.

\begin{figure}
\begin{center}
\includegraphics[scale=0.4,angle=0]{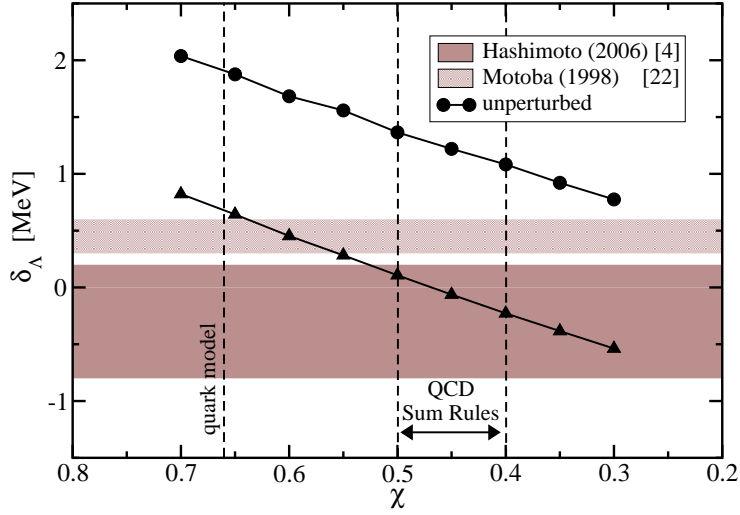}
\caption{\label{figB} 
Evolution of the spin-orbit splitting
$\delta_\Lambda = \epsilon_\Lambda (1p^{-1}_{1/2}) - 
\epsilon_\Lambda (1p^{-1}_{3/2})$ for $^{16}_\Lambda O$ as function of the ratio
$\chi$ between background scalar-vector self-energies of $\Lambda$ vs. nucleon.  The dashed line at $\chi = 2/3$ marks the simple quark model value. Also indicated is the $\chi$ interval suggested by a QCD sum rule analysis~\cite{Cohen:1994wm,Jin:1993fr}. Calculations with (without) the chiral SU(3) spin-orbit correction (\ref{so}) are denoted by triangles (circles).
The dark shaded area represents an estimate of $\delta_\Lambda$ ($-0.8$ MeV $\le \delta_\Lambda \le 0.2$ MeV) based on the measured excitation energy difference 
$\Delta E (0^+ -2^+)$ in
$~^{16}_\Lambda$O~\cite{Hashimoto_rev}, while the older determination is 
represented by a light shaded area ($0.3$ MeV $\le \delta_\Lambda \le 0.6$ MeV)
\cite{Motoba:1998td}. 
These estimates are consistent with recent results from hypernuclear 
$\gamma$ ray spectroscopy~\cite{Kohri:2001ncAjimura:2001naAkikawa:2002tm}.
}
\end{center}
\end{figure}

In Fig.~\ref{figB} we plot the $\Lambda$ spin-orbit splitting
$\delta_\Lambda = \epsilon_\Lambda (1p^{-1}_{1/2}) - 
\epsilon_\Lambda (1p^{-1}_{3/2})$ in $^{16}_\Lambda O$ 
as function of the ratio
$\chi$ between the scalar-vector background mean fields 
for the $\Lambda$ hyperon and for the nucleon. The filled 
circles show the spin-orbit splitting produced by the 
scalar-vector background fields alone. Even for 
unnaturally small values of $\chi$, the splitting 
remains systematically too large in comparison with 
the empirical bounds. Introducing the model-independent 
spin-orbit contribution from second order
pion exchange as in Eq.(\ref{eqn:ls}), the previous 
line is shifted downward (filled triangles) by 
about 1.3 MeV. 
With $\chi$, as suggested by QCD sum rules analysis~\cite{Cohen:1994wm,Jin:1993fr},
smaller than $2/3$,
the small or even vanishing spin-orbit splitting is now reproduced
in agreement with empirical estimates~\cite{Hashimoto_rev,Motoba:1998td}.
 
3. {\it Concluding remarks.} The compensating mechanism 
for the spin-orbit interaction of a $\Lambda$ in nuclear 
matter, proposed in Ref.~\cite{Kaiser:2004fe}, appears to
 be successful in explaining the very small spin-orbit 
splitting in finite $\Lambda$ hypernuclei. We emphasize 
again that this mechanism, driven by the second order 
pion exchange tensor force between $\Lambda$ and nucleon 
with  intermediate $\Sigma$, is model-independent in 
that it relies only on SU(3) chiral dynamics with 
empirically well known constants. This well controlled 
intermediate-range effect (independent of any 
regularization procedure) counteracts short-distance 
spin-orbit forces. The corresponding effect in 
ordinary nuclei is neutralized by three-body 
spin-orbit terms (induced by two-pion exchange with virtual
$\Delta$-isobar excitation)
which are absent in hypernuclei. 
Systematic applications of the present 
framework to larger classes of hypernuclei are under way. 

{\it Acknowledgements}. We thank A.~Gal for helpful 
discussions. This research was partly supported by 
BMBF, GSI, INFN, MURST, MZOS (project 1191005-1010) 
and by the DFG cluster of 
excellence Origin and Structure of the Universe. 


\begin{thebibliography}{00}





\bibitem{danysz}
M. Danysz and J. Pniewski, Phil. Mag. 44 (1953) 348.

\bibitem{hyper_review_1}
R.~E.~Chrien and C.~B.~Dover,
Ann.\ Rev.\ Nucl.\ Part.\ Sci.\  {\bf 39} (1989) 113;
D.~J.~Millener, C.~B.~Dover and A.~Gal,
Phys.\ Rev.\  C {\bf 38} (1988) 2700;
H.~Bando,
Prog.\ Theor.\ Phys.\  {\bf 81} (1985) 1.

\bibitem{hyper_review_2}
Proc. Int. Conf. HYP 2003, Nucl. Phys. A {\bf 754} (2005) 1-489; 
Proc. Int. Conf. HYP 2006, Eur. Phys. J. A {\bf 33} (2007) 243-301.

\bibitem{Hashimoto_rev}
O.~Hashimoto and H.~Tamura,
Prog.\ Part.\ Nucl.\ Phys.\  {\bf 57} (2006) 564, and references therein.

\bibitem{Kohri:2001ncAjimura:2001naAkikawa:2002tm}
H.~Kohri {\it et al.}  [AGS-E929 Collaboration],
Phys.\ Rev.\  C {\bf 65} (2002) 034607;
S.~Ajimura {\it et al.},
Phys.\ Rev.\ Lett.\  {\bf 86} (2001) 4255;
H.~Akikawa {\it et al.},
Phys.\ Rev.\ Lett.\  {\bf 88} (2002) 082501.


\bibitem{wei_bro}
R. Brockmann and W. Weise, Phys. Lett. B {\bf 69} (1977) 167; 
Nucl. Phys. A {\bf 355} (1981) 365. 


\bibitem{noblepirnerjenn}
H. J. Pirner, Phys. Lett. B {\bf 85} (1979) 190;
J. V. Noble, Phys. Lett. B {\bf 89} (1980) 325;
B. K. Jennings, Phys. Lett. B {\bf 246} (1990) 325.  

\bibitem{tensor}
J.~Mares and B.~K.~Jennings,
Phys.\ Rev.\ C {\bf 49} (1994) 2472.


\bibitem{Tsushima}
K.~Tsushima, K.~Saito, J.~Haidenbauer and A.~W.~Thomas,
Nucl.\ Phys.\ A {\bf 630} (1998) 691;
K.~Saito, K.~Tsushima and A.~W.~Thomas,
Prog.\ Part.\ Nucl.\ Phys.\  {\bf 58} (2007) 1.

\bibitem{Keil:1999hk}
C.~M.~Keil, F.~Hofmann and H.~Lenske,
Phys.\ Rev.\ C {\bf 61} (2000) 064309.


\bibitem{Kaiser:2001jx}
N.~Kaiser, S.~Fritsch and W.~Weise,
Nucl.\ Phys.\ A {\bf 697} (2002) 255.

\bibitem{Fritsch:2004nx}
S.~Fritsch, N.~Kaiser and W.~Weise,
Nucl.\ Phys.\ A {\bf 750} (2005) 259.


\bibitem{Kaiser:2004fe}
N.~Kaiser and W.~Weise,
Phys.\ Rev.\ C {\bf 71} (2005) 015203.


\bibitem{Finelli:2003fk}
P.~Finelli, N.~Kaiser, D.~Vretenar and W.~Weise,
Nucl.\ Phys.\ A {\bf 735} (2004) 449.

\bibitem{Finelli:2005ni}
P.~Finelli, N.~Kaiser, D.~Vretenar and W.~Weise,
Nucl.\ Phys.\ A {\bf 770} (2006) 1.

\bibitem{Finelli:2006xh}
P.~Finelli, N.~Kaiser, D.~Vretenar and W.~Weise,
Nucl. Phys. A {\bf 791} (2007) 57.


\bibitem{Camalich:2006is}
J.~M.~Camalich and M.~J.~V.~Vacas,
Phys.\ Rev.\  C {\bf 75} (2007) 035207.

\bibitem{SerotFurnstahl}
B.~D.~Serot and J.~D.~Walecka,
Int.\ J.\ Mod.\ Phys.\  E {\bf 6} (1997) 515;
R.~J.~Furnstahl,
Lect.\ Notes Phys.\  {\bf 641} (2004) 1.

\bibitem{Fuchs:1995as}
C.~Fuchs, H.~Lenske and H.~H.~Wolter,
Phys.\ Rev.\  C {\bf 52} (1995) 3043.

\bibitem{Cohen:1994wm}
T.~D.~Cohen, R.~J.~Furnstahl, D.~K.~Griegel and X.~m.~Jin,
Prog.\ Part.\ Nucl.\ Phys.\  {\bf 35} (1995) 221.

\bibitem{Jin:1993fr}
X.~m.~Jin and R.~J.~Furnstahl,
Phys.\ Rev.\ C {\bf 49} (1994) 1190.



\bibitem{Motoba:1998td}
T.~Motoba,
Nucl.\ Phys.\  A {\bf 639} (1998) 135.

\bibitem{Chabanat:1998}
E. Chabanat, P. Bonche, P. Haensel, J. Meyer and R. Schaeffer, 
Nucl. Phys. A  {\bf 635} (1998) 231.

\bibitem{KFW:2003}
N.~Kaiser, S.~Fritsch and W.~Weise,
Nucl. Phys. A  {\bf 724} (2003) 47.

\bibitem{Kaiser:2003}
N.~Kaiser, Phys. Rev.  C  {\bf 68} (2003) 054001.

\bibitem{Plohl:2006hy}
O.~Plohl and C.~Fuchs,
Phys.\ Rev.\  C {\bf 74} (2006) 034325.

\end{thebibliography}
\end{document}